  \documentclass[%
  reprint,
  superscriptaddress,
   amsmath,amssymb,
   aps,
  prc,
  showpacs
  ]{revtex4-2}
  \usepackage[dvipdfmx]{graphicx}
  \graphicspath{{./fig/}}  
  \usepackage{dcolumn}
  \usepackage{bm}
  \usepackage[T1]{fontenc}
  \usepackage{braket}
  \RequirePackage[normalem]{ulem}
  \usepackage{hyperref}
  \hypersetup{pdfpagemode={UseOutlines},
  bookmarksopen=true,
  bookmarksopenlevel=0,
  hypertexnames=false,
  colorlinks=true,
  citecolor=blue,
  linkcolor=blue,
  urlcolor=blue,
  allcolors=blue,
  pdfstartview={FitV},
  unicode,
  breaklinks=true,
  }
  
  \usepackage{etoolbox} 
\begin{document}

\title{
  $\alpha$ knockout reaction as a new probe for $\alpha$ formation in $\alpha$-decay nuclei
}

\author{Kazuki~Yoshida}
  \email[]{yoshida.kazuki@jaea.go.jp}
  \affiliation{Advanced Science Research Center, Japan Atomic Energy Agency,
  Tokai, Ibaraki 319-1195, Japan}
  
\author{Junki~Tanaka}
  \affiliation{RIKEN Nishina Center for Accelerator-Based Science,
   2-1 Hirosawa, Wako 351-0198, Japan}

\date{\today}

\begin{abstract}
\begin{description}

\item[Background]
  The $\alpha$-decay phenomenon has been studied for more than a century. 
  Its mechanism is simply explained by two factors, the $\alpha$ formation probability and the penetration process through the potential barrier.
\item[Purpose]
  As an alternative to the $\alpha$-decay lifetime measurement,
  we propose the proton-induced $\alpha$ knockout reaction, ($p$,$p\alpha$), as a new probe for the surface $\alpha$ formation probability of $\alpha$-decay nuclei.
\item[Method]
  The $^{210,212}$Po($p$,$p\alpha$)$^{206,208}$Pb reaction is described by the 
  distorted-wave impulse approximation framework.
\item[Results]
  It is shown that the $^{212}$Po/$^{210}$Po ratio of the  
  $\alpha$ knockout cross sections agrees with that of the surface
  $\alpha$ formation probabilities
  determined by lifetime measurements.
\item[Conclusions]
  It was confirmed that the ($p$,$p\alpha$) reaction cross sections correspond to the $\alpha$ formation probability of the nuclear surface.
  The result implies that the ($p$,$p\alpha$) cross section is a direct probe for the
  surface $\alpha$ formation probability, which is an essential quantity for complete understanding of the $\alpha$-decay phenomenon.
\end{description}

\end{abstract}


\maketitle

\section{Introduction}
$\alpha$ decay, the emission of $^{4}$He nuclei, was discovered by Rutherford~\cite{Rutherford1899,rutherford1903}, and theoretically explained by 
Gamow~\cite{Gamow1928}. According to the theory, intra-nuclear preformed $\alpha$ 
particles are released out of the nucleus at a certain half-life due to the quantum tunneling 
effect. 
However, from the viewpoint of nuclear structure theory, it is still challenging to quantitatively describe the $\alpha$ amplitude from nucleon degrees of freedom.

The $\alpha$-decay width $\Gamma_\alpha$ is given by a product of the penetrability of the Coulomb barrier $P$ and the reduced width $\gamma^2$~\cite{Thomas54,Lane58}:
\begin{align}
\Gamma_\alpha = 2P\gamma^2. 
\label{eq_gamma}
\end{align}
See also Eqs.~(\ref{eq_gamma2}) and (\ref{eq_halflife}).
The reduced width is the probability of the $\alpha$ cluster formation on the nuclear surface. 
Currently, the reduced width (the number of preformed $\alpha$ particles) is estimated for a wide range of atomic nuclei 
from the $\alpha$-decay half-life measurements and estimated penetrabilities~\cite{rasmussen1959alpha,QI2019214} (see also Fig. \ref{fig_result} and the experimental data~\cite{Andreyev13} therein).
However, it is easy to imagine that the actual structure of heavy nuclei is rather complicated. 
Theoretical estimations of the reduced width are model dependent, and their products with the penetrability are constrained by half-life measurements.
In theoretical $\alpha$-decay researches, it has long been pointed out that $\alpha$-core two-body description in which $\alpha$ particles 
have certain amplitude inside the nucleus are inconsistent with the saturation of the nucleus, and a wave function with amplitude only 
on the nucleus surface was suggested \cite{tonozuka79surface}. 
The same property is also shown by the calculation of the density distribution of $\alpha$ particles on the tin isotope surface 
by density functional theory \cite{typel14}. 
In the light mass region, theoretical studies of $\alpha$ clusters based on the nucleon degrees of freedom also showed that $\alpha$ 
amplitude in the internal region is suppressed and locally has a peak on the surface 
due to the antisymmetrization between nucleons. 
See Refs.~\cite{Horiuchi14,Chiba17} for recent results of the $\alpha$ amplitude of $^{16}$O and $^{20}$Ne for example. 
Such $\alpha$ amplitude is shown to give a consistent ($p$,$p\alpha$) knockout cross section with existing data{~\cite{Yoshida18}.
In the mass region heavier than the lead nucleus, the reduced widths of isotopes rapidly 
increase on the neutron excess side at the neutron magic number 126. For example, $^{212}$Po 
is composed of $\alpha$ particles and a double closed shell of $^{208}$Pb, and its reduced  
width is 10.6 times larger than that of $^{210}$Po~\cite{rasmussen1959alpha}, 
which has only two fewer neutrons (see also Fig.~\ref{fig_result}). 

Besides the $\alpha$-decay measurements, attempts from direct reaction approaches have been made to probe the $\alpha$ formation amplitude directly. 
Transfer reactions such as $\alpha$-particle pickup reaction \cite{arimatreatise} and stripping reaction~\cite{Devries75} have been used as alternative probes for measuring the surface $\alpha$-particle formation probability. Although these reactions are peripheral and suitable to probe the $\alpha$ amplitude on the nuclear surface, they are not free from the Coulomb barrier due to its relatively low reaction energies.  
Although a low energy beam is required to satisfy the momentum matching in the transfer reaction, the channel coupling may play some roles at such reaction  energy region, and extraction of the $\alpha$ amplitude by theoretical reaction analysis will be complicated.
 \begin{figure}[htbp]
    \centering
    \includegraphics[width=0.35\textwidth]{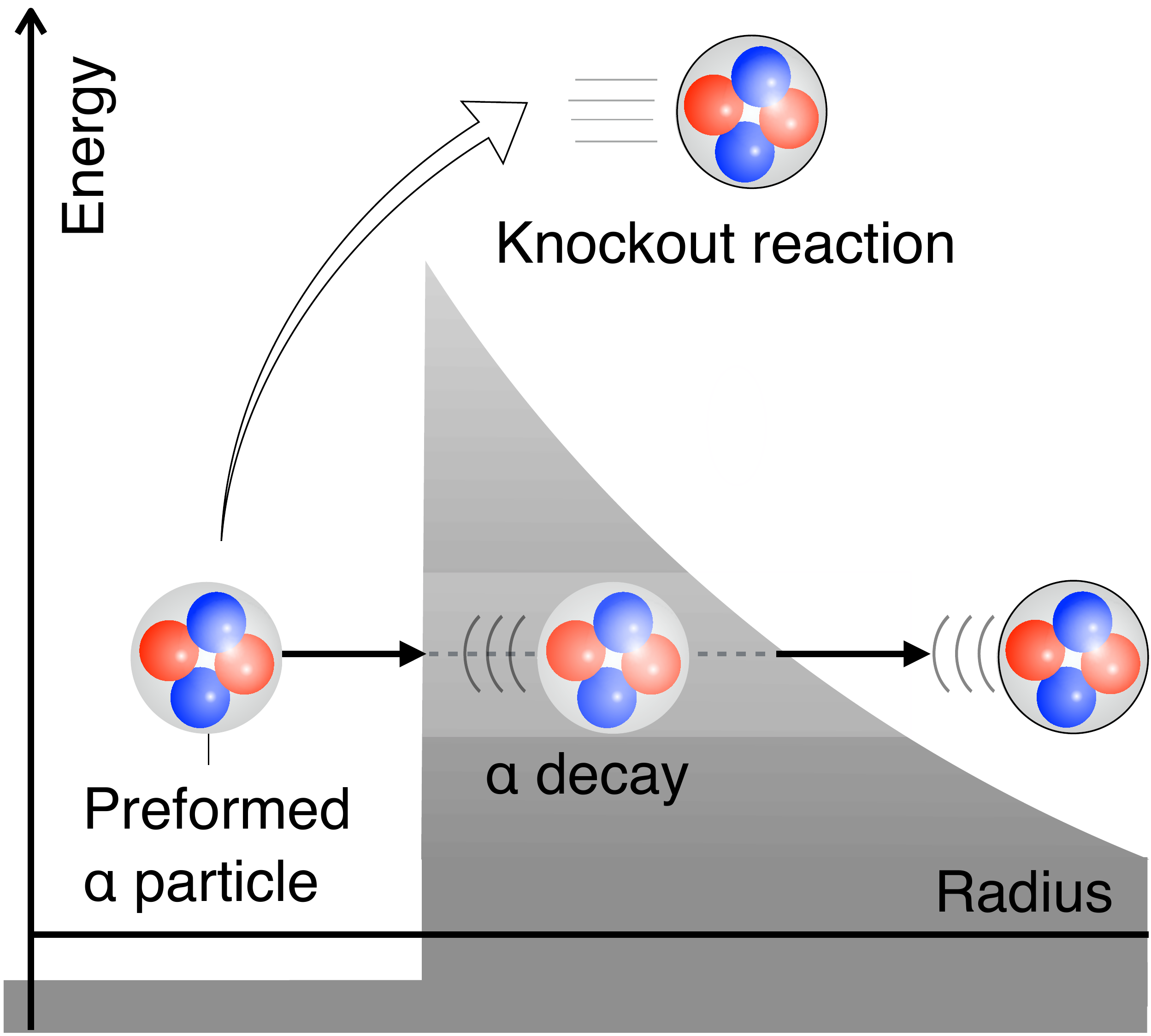}
    \caption{
     Contrast between $\alpha$ decay and knockout reaction. The preformed $\alpha$ 
     particle is tunneling through a
     Coulomb barrier in the decay process, while 
     the kinetic energy beyond the 
     Coulomb barrier is instantly given to the $\alpha$ particle in the knockout reaction and jumps over the barrier.
    }
    \label{fig_concept}
 \end{figure}
  A recent experimental study has shown that the $\alpha$ knockout reaction, 
  which knocks out $\alpha$ particles with high-energy protons, is useful to 
  observe the $\alpha$ formation in medium and heavy nuclear regions~\cite{Tanaka21}. 
  The $\alpha$ particle is emitted from a nucleus less affected by the Coulomb barrier because the high-energy proton knocks out the $\alpha$ particle far above the Coulomb barrier with large energy and momentum transfer.
  It is also shown that the reaction is less affected by the internal nuclear structure because of the surface sensitivity of the reaction~\cite{Yoshida16}.
  Figure \ref{fig_concept} shows the contrast between $\alpha$ decay and $\alpha$ knockout reaction. 
  It illustrates that the number of $\alpha$ particles is evaluated from a completely 
  different kinematics. The $\alpha$ particle emitted by the knockout reaction gets over 
  the Coulomb barrier and is less affected by its structure. 
  When this method is applied to $\alpha$-decay nuclei, it is necessary to connect 
  the measured cross section and the number of preformed $\alpha$ particles  
  using the reaction theory. 
  Because ($p$,$p\alpha$) data of heavy nuclei are very limited, 
  this paper assumes the 
  number of preformed $\alpha$ particles, the wave functions, from $\alpha$-decay studies 
  on $^{210,212}$Po.
  The reaction theory provides the relation between cross section and 
  the number of preformed $\alpha$ particles. 
  The knockout cross sections of $^{210,212}$Po($p$,$p\alpha$) are predicted 
  using the wave functions.
  
  In actual measurements, the reverse process gives the 
  number of preformed $\alpha$ particles from the observables of the $\alpha$ knockout reaction.

\section{Theoretical framework}
Based on the $R$-matrix theory of the cluster decay process,
    the $\alpha$-decay half-life $T_{1/2}$ and the decay width $\Gamma_\alpha$ are determined by the penetrability $P$ and the reduced width $\gamma^2$~\cite{Thomas54,Lane58}:
    \begin{align}
      \Gamma_\alpha
      &=
      2P(R)\times \gamma^2(R) 
      \nonumber \\
      &=
      2\frac{kR}{\left|H_l^{+}(\chi,\rho)\right|^2}
      \times
      \frac{\hbar^2}{2\mu R}|RF(R)|^2
      \nonumber \\
      &=
      \left| \frac{RF(R)}{H_l^+ (\chi,\rho)} \right|^2
      \frac{\hbar^2 k}{\mu},
      \label{eq_gamma2}
       \\
      T_{1/2} 
      &=
      \frac{\hbar \mathrm{ln} 2}{\Gamma_\alpha}
      \approx
      \frac{\mathrm{ln}2}{\nu}
      \left| \frac{H_l^+ (\chi,\rho)}{RF(R)} \right|^2,
      \label{eq_halflife}
    \end{align}
    where $k$ and $\nu = \hbar k /\mu$ are the outgoing wave number and velocity of the emitted $\alpha$, respectively, 
    $RF(R)$ is the surface $\alpha$ formation amplitude ($\alpha$ reduced width amplitude) 
    at the radius $R$ of the nuclear surface.
    $H_l^+$ is the Coulomb-Hankel function with an angular momentum $l$.
    Its arguments $\rho$ and $\chi$ are defined in the usual manner; see Ref.~\cite{Qi09_prl}
    for details.
    Following the prior researches of the $\alpha$ knockout reaction,
    the distorted wave impulse approximation 
    framework (DWIA)
    is employed to describe the Po($p$,$p\alpha$)Pb cross section.
    Details of the theoretical framework can be found in 
    Refs.~\cite{Yoshida16,Lyu18,Yoshida18,Lyu19,Yoshida19,Taniguchi_48Ti}.
    Since the DWIA descriptions of the ($p$,$p\alpha)$ and ($p$,$pN$) reactions are essentially the same, 
    it will be also helpful to refer to the DWIA description of the
    ($p$,$pN$) reaction in Sec.~3 of a recent review article~\cite{Wakasa17}.

    Leaving the details to the above-mentioned references,      
    the DWIA framework for this study is briefly introduced as follows.
    The incident and emitted protons are labeled as particles 0 and 1, respectively.
    The momentum (wave number) and its solid angle and the total energy of
    particle $i = 0, 1, \alpha$ are
    denoted by $\bm{K}_i$, $\Omega_i$, $E_i$, respectively.
    The triple differential cross section (TDX) of the A($p$,$p\alpha$)B reaction is given by
    \begin{align}
      \frac{d^3\sigma}
      {dE_1^\mathrm{A} d\Omega_1^\mathrm{A} d\Omega_\alpha^\mathrm{A}
      }
      =&
      F_{\mathrm{kin}}^{\mathrm{A}}
      \frac{E_1 E_2 E_{\mathrm{B}}}{E_1^{\mathrm{A}} E_2^{\mathrm{A}} E_{\mathrm{B}}^{\mathrm{A}}}
      \frac{(2\pi)^4}{\hbar v_\alpha}
      \frac{1}{2l+1} \nonumber \\
      &\times
      \frac{(2\pi \hbar^2)^2}{\mu_{p\alpha}}
      \frac{d\sigma_{p\alpha}}{d\Omega_{p\alpha}}
      \left|
      \bar{T}
      \right| ^2.
    \end{align}
    The total energy of the emitted proton and the emission directions of the proton 
    and $\alpha$ are denoted by $E_1^{\mathrm{A}}$, $\Omega_1^{\mathrm{A}}$, and
    $\Omega_2^{\mathrm{A}}$, respectively.
    Note that the Po($p$,$p\alpha$)Pb reaction in inverse kinematics is
    considered in this study;
    the Po beam is bombarded on the proton target.
    Quantities with (without) superscript A are evaluated in the projectile 
    rest frame
    (center-of-mass frame).
    $v_\alpha$ is the relative velocity of $p$ and $^{210,212}$Po in the initial state,
    and $\mu_{p\alpha}$ is the reduced mass of these particles.
    $l$ is the orbital angular momentum of the $\alpha$--Pb cluster state;
    $l=0$ is assumed in this study.
    The kinematical factor $F_\mathrm{kin}^\mathrm{A}$ is defined by
    \begin{align}
      F_\mathrm{kin}^\mathrm{A}
      &=
      \frac{E_1^\mathrm{A} K_1^\mathrm{A} E_2^\mathrm{A} K_2^\mathrm{A}}{(\hbar c)^4} \nonumber \\
       &\times
       \left[
        1
        + \frac{E_2^\mathrm{A}}{E_\mathrm{B}^\mathrm{A}}
        + \frac{E_2^\mathrm{A}}{E_\mathrm{B}^\mathrm{A}}
          \frac{\left(\bm{K}_1^\mathrm{A} - \bm{K}_0^\mathrm{A} - \bm{K}_\mathrm{A}^\mathrm{A}\right)\cdot \bm{K}_2^\mathrm{A}}
          {\left(K_2^\mathrm{A}\right)^2}
      \right]^{-1}.
    \end{align}
    Note that $\bm{K}_\mathrm{A}^\mathrm{A} = 0$ in this case.
    $d\sigma_{p\alpha}/d\Omega_{p\alpha}$ is the $p$-$\alpha$ differential 
    cross section.
    The reduced transition matrix $\bar{T}$ is given by
    \begin{align}
      \bar{T}
      &=
      \int 
      \bar{\chi}(\bm{R}) F(\bm{R})
      \,d\bm{R}, 
      \label{eq_tmat}
      \\
      \bar{\chi}(\bm{R})
      &\equiv
      \chi_{1}^{*(-)}(\bm{R}) \chi_{\alpha}^{*(-)}(\bm{R}) \chi_{0}^{(+)}(\bm{R})
      e^{-\bm{K}_0\cdot\bm{R}A_\alpha/A},
    \end{align}
    where $\chi_{i}$ ($i = 0, 1, \alpha$) are the distorted waves between $p$-A, $p$-B, $\alpha$-B,
    respectively.
    $A_\alpha$ and $A$ are the mass numbers of $\alpha$ and $^{210,212}$Po.
    The $\alpha$ cluster wave function of $\alpha$-B system is denoted by $F(\bm{R})$,
    and its radial part is the $\alpha$ formation amplitude $F(R)$.
    It should be noted here that the $\alpha$ spectroscopic factor ($S_\alpha$)
    is implicitly 
    taken into account in the squared norm of $F(\bm{R})$ in the present formalism.

    The transition matrix density (TMD)~\cite{Wakasa17} is a good measure of the peripherality of the reaction.
    It is defined by
    \begin{align}
      \delta(R)
      &=
      {\bar{T}}^{*}
      \int \bar{\chi}(\bm{R})F(\bm{R}) R^2 d\hat{\bm{R}}.
      \label{eq_tmd}
    \end{align}
    Equations~(\ref{eq_tmat}) and (\ref{eq_tmd}) lead to 
    \begin{align}
      \int \mathrm{Re}\left[ \delta(R) \right] dR &= \left| \bar{T} \right|^2.
    \end{align}
    Therefore, $\mathrm{Re}\left[ \delta(R) \right]$ can be regarded as 
    a radial distribution of the $|\bar{T}|^2$ and therefore the cross section.
    This quantity is discussed in Sec.~\ref{subsec_TDX}
    to investigate the peripherality of the reaction.

    Considering the experimental setup in the inverse kinematics,
    the momentum distribution of the residue B is discussed in the following. 
    It is obtained from TDX as
    \begin{align}
      \frac{d\sigma}{d\bm{K}_\mathrm{B}^\mathrm{A}}
      =
      \int  
      &\delta\left( E_f - E_i\right) 
      \delta\left(\bm{K}_f - \bm{K}_i\right) \nonumber \\
      &\times\frac{d^3\sigma}{dE_1^\mathrm{A}d\Omega_1^\mathrm{A}d\Omega_\alpha^\mathrm{A}}
      d\bm{K}_1^\mathrm{A} d\bm{K}_\alpha^\mathrm{A}.
    \end{align}
    Two $\delta$ functions are inserted to ensure the energy and momentum conservation.
    The one-dimensional longitudinal momentum distribution (LMD) is then 
    defined by
    \begin{align}
      \frac{d\sigma}{dK_{\mathrm{B}z}^\mathrm{A}}
      &=
      2\pi
      \int \frac{d\sigma}{d\bm{K}_\mathrm{B}^\mathrm{A}} 
      K_{\mathrm{B}b}^\mathrm{A} d K_{\mathrm{B}b}^\mathrm{A}.
      \label{eq_lmd}
    \end{align}
    The axial distance and axial coordinate of $\bm{K}_\mathrm{B}^\mathrm{A}$
    in the cylindrical coordinates are denoted by $K_\mathrm{Bb}^\mathrm{A}$
    and $K_\mathrm{Bz}^\mathrm{A}$, respectively.
    The total cross section $\sigma$ is obtained by integrating 
    Eq.~(\ref{eq_lmd}) over $K_{\mathrm{B}z}^\mathrm{A}$.

\section{Result and discussion}
  \subsection{Numerical inputs}
    $^{210,212}$Po beams at 200 MeV/nucleon 
    are considered in this study.
    The global optical potential parametrization of the
    proton-nucleus scattering by Koning and Delaroche~\cite{Koning03}
    is adopted to describe the $p$-A and $p$-B distorted waves.
    For the $\alpha$-B distorted wave, optical potential proposed by
    Avrigeanu, Hodgson, and Avrigeanu~\cite{Avrigeanu94} is employed.
    This is an extension to lower energies of the former work by Nolte \textit{et al}.~\cite{Nolte87}.
    The Melbourne $g$-matrix $NN$ interaction~\cite{Amos00} is applied to the folding model~\cite{Toyokawa13}
    to obtain the $p$-$\alpha$ effective interaction.
    Using this effective interaction,
    $d\sigma_{p\alpha}/d\Omega_{p\alpha}$ at required $p$-$\alpha$ scattering
    energy and angle are calculated in the DWIA calculations.

    Regarding the $\alpha$-$^{206,208}$Pb cluster wave function,
    we employ the $\alpha$-formation amplitude by Qi~\cite{Qi2010}
    shown in Fig.~\ref{fig_amp_tmd}(a).
    This wave functions has locally peaked amplitude on the surface and suppressed amplitude inside.
    These amplitudes are proposed to investigate the 
    sudden and significant 
    suppression of the $\alpha$ formation and
    decay width of $^{210}$Po compared to $^{212}$Po due to $N=126$ magicity.
    
    \begin{figure}[htbp]
    \centering
    \includegraphics[width=0.40\textwidth]{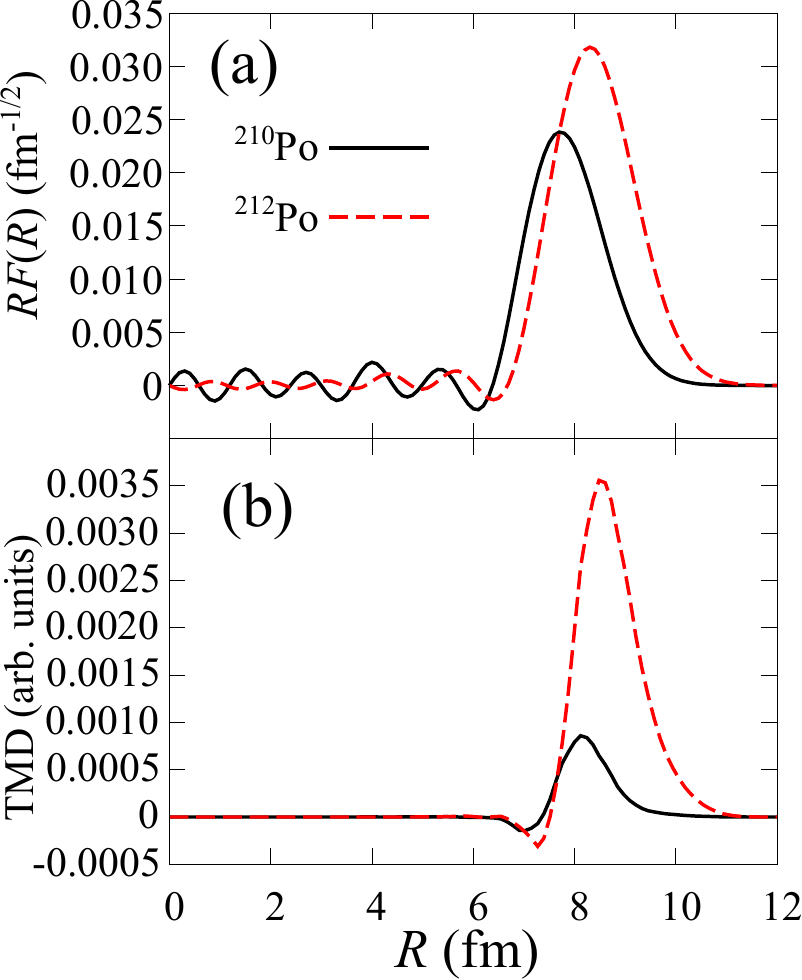}
    \caption{
      (a) 
      $\alpha$-$^{206,208}$Pb amplitude of $^{210,212}$Po taken
      from Fig.~3 of Ref.~\cite{Qi2010}.
      (b)
      TMD of $^{210,212}$Po($p$,$p\alpha$)$^{206,208}$Pb reaction
      given in arbitrary units.
    }
    \label{fig_amp_tmd}
    \end{figure}

    The reaction kinematics is fixed as follows to realize the 
    recoil-less condition, i.e., the residue is at rest in the Po rest frame.
    The reaction is coplanar, $\phi_{1}^\mathrm{A} = 0^\circ$
    and $\phi_{\alpha}^\mathrm{A} = 180^\circ$.
    The emission angle of $p$ is fixed at 
    $\theta_{1}^\mathrm{A} = 114.0^\circ$ for both 
    $^{210}$Po and $^{212}$Po cases.
    As for the emitted $\alpha$,
    $\theta_{\alpha}^\mathrm{A} = 130.0^\circ$ and $129.6^\circ$
    for $^{210}$Po and $^{212}$Po cases, respectively.
    The proton emission energy $T_1^\mathrm{A}$ is explicitly varied 
    around $T_1^\mathrm{A} = 140$~MeV and the others
    are determined by the conservation law.
    Note that these quantities are given in the projectile (Po) rest frame.

  \subsection{Po($p$,$p\alpha$)Pb cross section}
  \label{subsec_TDX}
    As shown in Fig.~\ref{fig_amp_tmd}(a),
    only a two-neutron-number difference in Po significantly pushes the 
    $\alpha$ amplitude outwards due to the $N=82$ magicity. 
    Such difference is magnified in 
    TDXs of $^{210,212}$Po($p$,$p\alpha$)$^{206,208}$Pb as shown
    in Fig.~\ref{fig_tdx};
    \begin{figure}[htbp]
    \centering
    \includegraphics[width=0.40\textwidth]{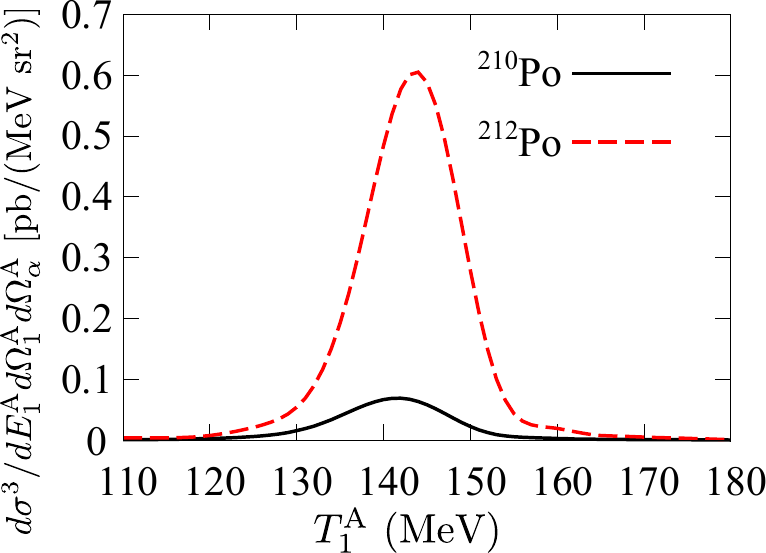}
    \caption{
    Proton emission energy distributions of the TDXs in $^{210,212}$Po($p$,$p\alpha$)$^{206,208}$Pb reactions.}
    \label{fig_tdx}
    \end{figure}
    The ratio of their peak heights is $8.79$, which
    is much larger than the ratio of their $S_\alpha$, 1.92.
    They are tabulated in Table~\ref{table_ratio}
    together with the LMD peak heights and the total cross sections.
    \begin{ruledtabular}
    \begin{table}
      \caption{
        The $\alpha$ spectroscopic factors $S_\alpha$,
        surface $\alpha$ formation probabilities $|RF(R)|^2$ at $R = 9.0$~fm,
        peak values of TDXs, LMDs, and the total cross sections
        of $^{210,212}$Po($p$,$p\alpha$)$^{206,208}$Pb.
        Cross sections are given in units of pb/(MeV sr$^2$), pb/(MeV/c), and mb,
        respectively.
      }
      \begin{tabular}{cccc}
                        &$^{210}$Po             & $^{212}$Po      & $^{212}$Po/$^{210}$Po        \\ \hline
        $S_\alpha$      & 7.63$\times 10^{-4}$ & 1.46$\times 10^{-3}$    & 1.92      \\
        $|RF(R)|^2$     & $5.20\times 10^{-5}$ & $5.28\times 10^{-4}$    & 10.2      \\
        TDX             & 0.069                & 0.605                   & 8.79      \\
        LMD             & 0.104                & 1.264                   & 12.2      \\
        total           & 0.0306               & 0.3644                  & 11.9      \\
      \end{tabular}
    \label{table_ratio}
    \end{table}
    \end{ruledtabular}
    Because of the surface sensitivity, 
    the ($p$,$p\alpha$) reaction probes the surface $\alpha$ formation probability 
    instead of the whole region ($S_\alpha$).
    It is also confirmed by TMDs of the reaction at the recoil-less
    condition, as shown in Fig.~\ref{fig_amp_tmd}(b).
    The $\alpha$ amplitude for $R \lesssim 8$~fm is strongly suppressed 
    by the absorption effect
    and does not contribute to the cross section.

    The LMDs of 
    $^{210,212}$Po($p$,$p\alpha$)$^{206,208}$Pb are shown in 
    Fig.~\ref{fig_lmd}.
    \begin{figure}[htbp]
      \centering
      \includegraphics[width=0.40\textwidth]{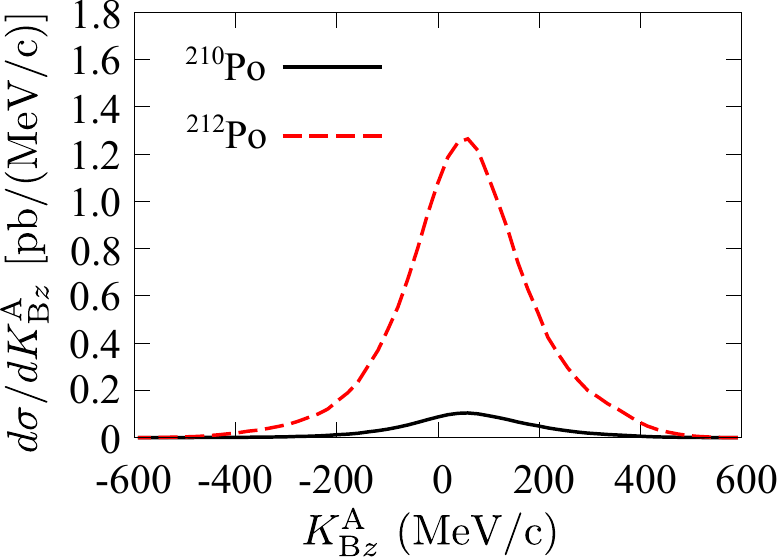}
      \caption{
        Longitudinal momentum distributions of the residues in $^{210,212}$Po($p$,$p\alpha$)$^{206,208}$Pb reactions.
      }
      \label{fig_lmd}
    \end{figure}
    Their peak heights are 1.264 and 0.104 pb/(MeV/c), 
    and their integrated values are 0.364 and 0.031~$\mu$b, respectively.
    As tabulated in Table~\ref{table_ratio}, the $^{212}$Po/$^{210}$Po ratios 
    of the LMD peak height and the total cross section agree well;
    they are 12.2 and 11.9, respectively.
    These values are in good agreement with the ratio of the surface $\alpha$ 
    formation probability
    from the structure theory [Ref.~\cite{Qi2010} and Fig.~\ref{fig_amp_tmd}(a)], 10.2.
    The shapes of the LMDs are asymmetric, and the peak positions are off from the center ($K_{\mathrm{B}z}^\mathrm{A} = 0$ MeV/c) to the positive side. It is due 
    to the strong Coulomb interaction between Pb and $\alpha$ in the final state. Because the 
    momentum of the emitted $\alpha$ in the projectile rest frame has a negative $z$ component, 
    i.e., $K_{\alpha z}^\mathrm{A} < 0$, the repulsive Coulomb force accelerates the residue to 
    the positive $z$ direction. Note the different behavior from the 
    knockout reactions in the region of light to medium mass nuclei. 
    See Fig.~3 of Ref.~\cite{Chen19} and Fig.~2 of \cite{Sun20} for examples of experimental data.
    In the light mass region, it has been pointed out in Ref.~\cite{Ogata15} that the reasons for the asymmetric shape 
    in LMDs are the phase volume effect and the distortion in scattering waves.
    In short, the former is because of less possibility of satisfying the
    energy and momentum conservation on the positive side of LMD.
    The latter is due to the final state attractive interaction between the
    residue and the emitted particles.
    In the present case, the repulsive Coulomb force overwhelms the attractive nuclear forces and 
    the phase volume effect due to its large $(Z_{A}-2)Z_{\alpha}$ values.

  As shown in Eq.~(\ref{eq_halflife}), 
  the surface $\alpha$ formation probability 
  $\left| RF(R) \right|^2$ is relevant to $\alpha$-decay lifetime.  
  Its isotopic trend among the polonium isotopes is extracted from the systematic 
  half-life measurements~\cite{Andreyev13}.
  \begin{figure}[htpb]
     \centering
     \includegraphics[width=0.45\textwidth]{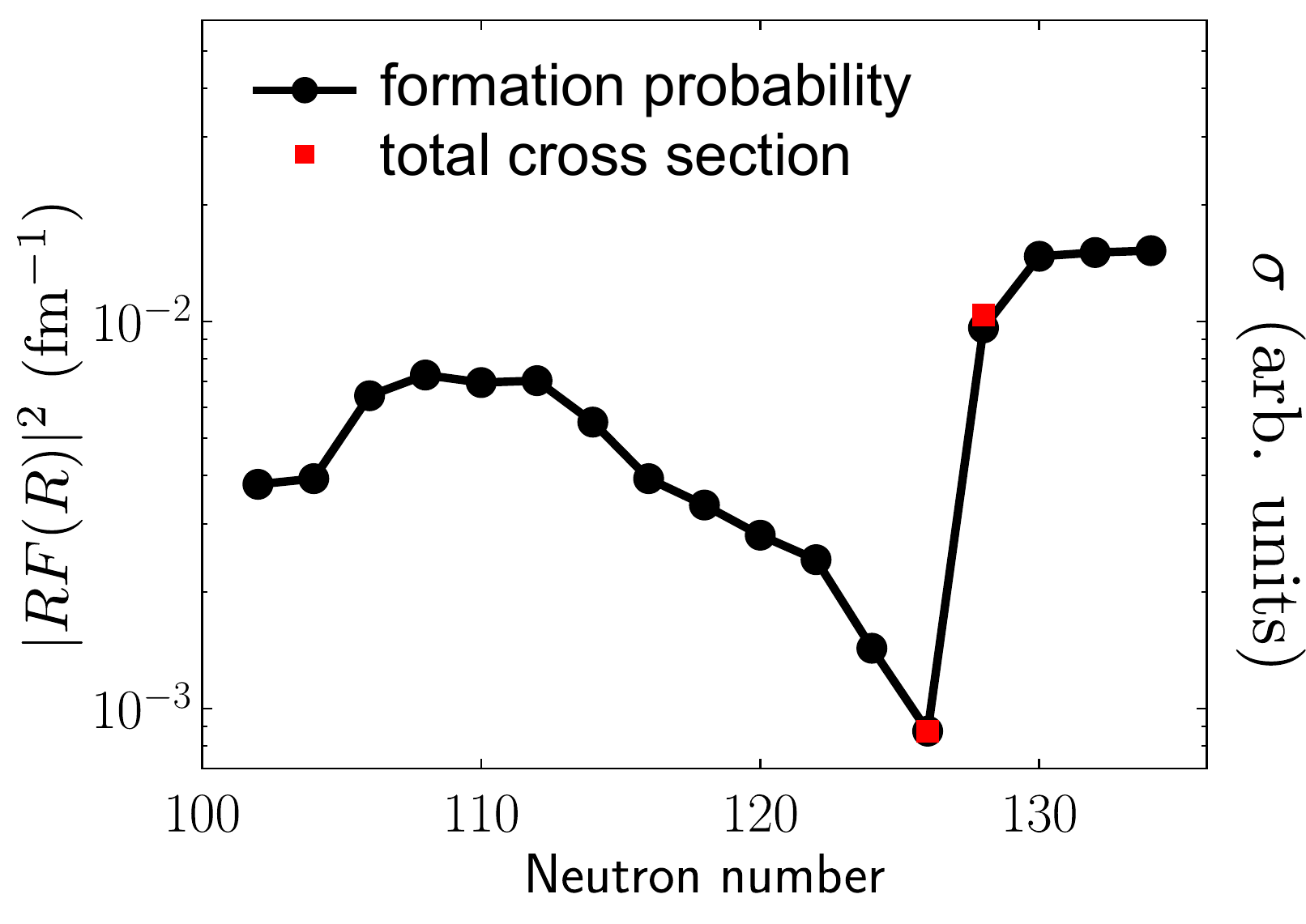}
     \caption{
       Comparison between surface $\alpha$ formation probabilities $|RF(R)|^2$
       of Po isotopes 
       extracted from experimental half-lives (circle)
       with the total knockout cross sections $\sigma$ of the present study (square).
       The data of formation probabilities are taken from Fig.~1(a) of Ref.~\cite{Andreyev13}.
       The total knockout cross sections in arbitrary units are normalized to the data at $^{210}$Po.}
     \label{fig_result}
  \end{figure}
  Because the total knockout cross section will be the most standard observable measured
  in experiments with high statistics,
  the trend of the total $\alpha$ knockout cross sections of 
  $^{210,212}$Po from the present theoretical study
  are compared with the data in Fig.~\ref{fig_result}.
  The total cross sections in arbitrary units are normalized to the
  $\alpha$ formation probability of $^{210}$Po.
  This result shows that the $^{212}$Po/$^{210}$Po ratio of the $\alpha$ knockout 
  cross sections agrees well with the ratio of their $\alpha$ formation probabilities.
  Ratios of the peak heights of the TDXs and LMDs give similar agreement, as shown
  in Table~\ref{table_ratio}.

  This agreement is accomplished by the surface sensitivity of the ($p$,$p\alpha$) reaction.
  The $^{210}$Po and $^{212}$Po were optimal for the first theoretical studies 
  because reliable wave functions were already prepared to reproduce the 
  $\alpha$-decay in these nuclei. 
  We expect that this relation holds in a wide range of $\alpha$ decay nuclei and therefore the ($p$,$p\alpha$) cross section will be a good measure  
  for the surface $\alpha$ formation probability in general.

\section{Summary}
We propose the ($p$,$p\alpha$) reaction as an alternative probe for the 
$\alpha$ formation probability of $\alpha$-decay nuclei.
The $^{210,212}$Po($p$,$p\alpha$)$^{206,208}$Pb reaction is described by 
the DWIA framework and  LMD and the total cross sections were obtained.
Their $^{212}$Po/$^{210}$Po ratios are
in good agreement with that of the $\alpha$ formation probability of  
the structure theory~\cite{Qi09_prl} and the experimental data~\cite{Andreyev13}.
This is because of the surface sensitivity of the ($p$,$p\alpha$) reaction---not $S_\alpha$
but the surface $\alpha$ formation probability,
which is also relevant to the $\alpha$-decay phenomena---is determined by the ($p$,$p\alpha$) cross section.
It was confirmed that the magnitudes of the ($p$,$p\alpha$) reaction cross sections correspond to the reduced width, which is the $\alpha$ formation probability on the nuclear surface.
Finally, experimental studies of the ($p$,$p\alpha$) reactions from $\alpha$-decay nuclei are eagerly awaited for a complete understanding of the long-standing $\alpha$-decay study.

\section*{ACKNOWLEDGMENTS}
  K. Y. thanks C. Qi for providing us with the $\alpha$ amplitude data.
  The authors thank T. Uesaka, S. Typel, and K. Ogata for 
  fruitful discussions.
  This work was supported by JSPS KAKENHI Grant No. JP20K14475.

\bibliography{ref}

\end{document}